\documentstyle [12pt]{article}

\def\K{{\cal K}}

\def\R{{\rm I\hspace{-.15em}R}}

\def\b{\begin{equation}} \def\e{\end{equation}}
\def\bd{\begin{displaystyle}} \def\ed{\end{displaystyle}}
\def\ba{\begin{array}}
\def\ea{\end{array}}

\def\bee{\begin{enumerate}}
\def\eee{\end{enumerate}}

\def\bes{\begin{eqnarray*}}
\def\ees{\end{eqnarray*}}
\def\be{\begin{eqnarray}}
\def\ee{\end{eqnarray}}

\title{Conformal linear gravity in de Sitter space}

\author{M.V. Takook$^{1,2}$\thanks{e-mail: takook@razi.ac.ir},\, M.R. Tanhayi$^3$\thanks{e-mail:
m$_{-}$tanhayi@iauctb.ac.ir} and S. Fatemi$^{4}$}
\date{\today}

\begin{document}

\maketitle  \centerline{\it $^1$ Department of Physics, Razi
University, Kermanshah, Iran} \centerline{\it $^2$ Groupe de
physique des particules, Universit\'e de
Montr\'eal,}\centerline{\it C.P. 6128, succ. centre-ville,
Montr\'eal, Qu\'ebec, Canada H3C 3J7} \centerline{\it $^3$
Department of Physics, Islamic Azad University, Central Tehran
Branch, Tehran, Iran }\centerline{\it $^4$ Science and research
branch, Islamic Azad University, Tehran, Iran}

\begin{abstract}
It has been shown that the theory of linear conformal quantum
gravity must include a tensor field of rank-3 and mixed symmetry
\cite{co5}. In this paper, we obtain the corresponding field
equation in de Sitter space. Then, in order to relate this field
with the symmetric tensor field of rank-2, $\K_{\alpha\beta}$
related to graviton, we will define homomorphisms between them.
Our main result is that if one insists $\K_{\alpha\beta}$ to be a
unitary irreducible representation of de Sitter and conformal
groups it must satisfy a filed equation of order 6, which is
obtained.

\end{abstract}

\vspace{0.5cm} {\it PACS numbers}: 04.62.+v, 98.80.Cq, 12.10.Dm
\vspace{0.5cm}


\setcounter{equation}{0}
\section{Introduction}

Gravitational fields are long range and seems to travel with the
speed of light, in the first approximation, at least, their
equations are expected to be conformally invariant (CI).
Einstein's theory of gravitation, in the background field method
($g_{\mu\nu}=g_{\mu\nu}^{BG}+h_{\mu\nu}$) and linear
approximation, can be considered as a theory of massless symmetric
tensor field of rank-2, $h_{\mu\nu}$ on a fixed background
$g_{\mu\nu}^{BG}$, such as de Sitter space. It is well known that
the massless fields propagate on the light-cone and are invariant
under the conformal group $SO(2,4)$. For spin $s\geq 1$ they are
invariant under the gauge transformation as well.

On the other hand, Einstein's theory of gravity seems perfect as a
classical theory. Experimental data have confirmed it and have
ruled out several possible alternatives. However, as a quantum
theory it is less satisfactory since, as soon as one couples to
matter, the first order quantum corrections lead to a divergent
$S$ matrix. These divergences are nonrenormalizable. Einstein's
classical theory of gravitation, as well as equation of
$h_{\mu\nu}$ is not CI thus could not be considered as a
comprehensive universal theory of gravitational fields. \\  In de
Sitter (dS) space, mass is not an invariant parameter for the set
of observable transformations under the dS group $SO(1,4)$.
Concept of light-cone propagation, however, does exist and leads
to the conformal invariance. ``Massless" is used in reference to
propagation on the dS light-cone (conformal invariance). The term
``massive" is refereed to fields that in their Minkowskian limit
(zero curvature) reduce to massive Minkowskian fields \cite{ms9}.
The conformal invariance, and the light-cone propagation,
constitutes the basis for constructing ``massless'' field in dS
space.

In previous papers, we used Dirac's six cone formalism to obtain
CI equations for the scalar, vector \cite{me}, and rank-2
symmetric tensor \cite{derotata} fields which transformed
according to the unitary irreducible representation (UIR) of dS
group. The conformal space and six-cone formalism was first used
by Dirac to obtain the CI equations \cite{dir}. This formalism
developed by Mack and Salam \cite{s6} and many others \cite{o7}.
This approach to conformal symmetry leads to the best path to
exploit the physical symmetry in contrast to approaches based on
group theoretical treatment of state vector spaces. This is
essentially because in the latter approach it would be much more
difficult to see how to break the symmetry down to Poincar\'e
invariance \cite{s6}.

Barut and B\"{o}hm \cite{ms9} have shown that for the physical
representation of the conformal group (UIR), the value of the
conformal Casimir operator is $9$. But according to calculation of
Binegar et al \cite{co5} for the tensor field of rank-2 and
conformal degree 0, this value becomes 8. Therefore tensor field
of rank-2 does not correspond to any UIR of the conformal group.
In other words, the tensor field that carries physical
representations of the conformal group must be a tensor field of
higher rank.\\ In this paper we propose and study a mixed symmetry
tensor field of rank-3, $\Psi_{abc}$, with conformal degree zero,
which transforms according to the UIR of the conformal group
\cite{co5,frhe,none}. By mixed symmetry we mean $$
\Psi_{abc}=-\Psi_{bac},\,\,\,\,\sum_{cycl}\Psi_{abc}=0,$$ while a
field of conformal degree zero satisfies
$u^d\partial_d\Psi_{abc}=0,\,\,a,b,c,d\equiv 0,1,...,5$, where
$u^d$ are the coordinates in $\R^{6}$. We then project this field
to dS space and define homomorphisms between the projected field,
$F_{\alpha\beta\gamma}$, and rank-2 symmetric tensor field
$\K_{\alpha\beta}$ on dS space ($\alpha,\beta\equiv 0,1,...,4$).
It has been shown that if one insists $\K_{\alpha\beta}$ to
transform according to the UIRs of dS and conformal groups it must
satisfy a field equation of order 6.

The paper is organized as follows. Section $2$ is devoted to a
brief review of the notations. In this section we recall Dirac's
manifestly covariant formalism of mixed symmetry tensor fields on
the six-cone and their projection to de Sitter space. Section 3
introduces CI wave equation with the subsidiary conditions
\emph{i.e.}, transversality and divergencelessness. Section $4$ is
devoted to define homomorphisms between $F_{\alpha\beta\gamma}$
and $\K_{\alpha\beta}$ on de Sitter space. Finally a brief
conclusion and an outlook for further investigation has been
presented.

\setcounter{equation}{0}
\section{Notation}

The dS metric is a solution of the cosmological Einstein's
equation with positive constant $\Lambda$. Recent astrophysical
data indicate that our universe might currently be in a dS phase
\cite{ja}. The importance of dS space has been primarily ignited
by the study of the inflationary model of the universe and quantum
gravity \cite{ja2}. The de Sitter space is identical to four
dimensional one-sheeted hyperboloid (intrinsic) embedded in five
dimensional flat space (ambient)
$$ X_H=\{x \in \R^5 ;x^2=\eta_{\alpha\beta} x^\alpha
 x^\beta =-H^{-2}\},\;\; \alpha,\beta=0,1,2,3,4, $$
where $\eta_{\alpha\beta}=$diag$(1,-1,-1,-1,-1)$ and $H$ is the
Hubble parameter.

The concept of conformal space was used by Dirac \cite{dir} to
demonstrate the field equations for spinor and vector fields in
$1+3$ dimensional space-time in manifestly CI form. The conformal
group $SO(2,4)$ acts nonlinearly on Minkowski coordinates. Dirac
proposed a manifestly conformally covariant formulation in which
the Minkowski coordinates are replaced by coordinates on which
$SO(2,4)$ acts linearly. The resulting theory is then formulated
on a 5 dimensional hypercone (named Dirac's six-cone) in a 6
dimensional space. Dirac's six-cone, or Dirac's projection cone,
is defined by \b u^2\equiv (u^0)^{2}-\vec
u^{2}+(u^5)^{2}=\eta_{ab} u^a u^b=0 ,\;\;
\eta_{ab}=\mbox{diag}(1,-1,-1,-1,-1,1),\e where $ \;u^{a} \in
\R^{6},$ and  $ \vec u \equiv(u^{1},u^{2},u^{3},u^{4})$. Reduction
to four dimensional (physical space-time) is achieved by
projection, that is by fixing the degrees of homogeneity of all
fields. Wave equations, subsidiary conditions, etc., must be
expressed in terms of operators that are defined intrinsically on
the cone. These are well-defined operators that map tensor fields
to tensor fields with the same rank on cone $u^2=0$ \cite{me,fr}.

We consider tensors of specific symmetry type that are transverse,
divergenceless and traceless,
\begin{enumerate}
\item[a)]{transversality}, $u_a\Psi^{ab\cdot\cdot\cdot}=0 ,$
\item[b)]{tracelessness}, $\Psi_{ab\cdot\cdot\cdot}^a=0 ,$
\item[c)]{divergencelessness}, $Grad_a\Psi^{ab\cdot\cdot\cdot}=0
,$ where the operator $Grad_a$ unlike $\partial_{a}$ is intrinsic
on the cone, and is defined by \cite{fr}:\b Grad_a\equiv
u_a\partial_b\partial^b-(2\hat{N_5}+4)\partial_a .\e
\end{enumerate} The action of second order
Casimir operator of conformal group on $\Psi$ is
\cite{co5,ms9,fr}:
$$ {\cal Q}_2\Psi^{cd\cdot\cdot\cdot}=\frac{1}{2}
L_{ab}L^{ab}\Psi^{cd\cdot\cdot\cdot}$$ \b =\left(-u^2
\partial^2+\hat{N_5}(\hat{N_5}+4)-2N+n_1(n_1+4)+n_2(n_2+2)+n_3^2\right)\Psi^{cd\cdot\cdot\cdot},\e where $n_1
\geq n_2\geq n_3\geq 0$ are integers that label the symmetry type
according to the lengths of the rows of the Yang diagrams and
$L_{ab}$ are the generators of the conformal Lie algebra. $\Psi$
is a tensor field of a definite rank and a definite symmetry. $N$
is the rank of the tensor field $\Psi^{abc\cdot\cdot}$ and
$\hat{N_5}$ is the conformal-degree operator defined by:\b
\hat{N_5}\equiv u^{a}\partial_{a}.\e On the cone $(u^2=0)$, the
second order Casimir operator of the conformal group, ${\cal Q}_2
$, reduces to a constant. Therefore it is not a suitable operator
to define CI wave equations. For example, for rank-2 symmetric
tensor field $\Psi^{cd}$, we have
$${\cal
Q}_2\Psi^{cd}=\left(\hat{N_5}(\hat{N_5}+4)+8\right)\Psi^{cd},$$
and for a mixed symmetry rank-3 tensor field $\Psi^{abc}$ we have
$${\cal
Q}_2\Psi^{abc}=\left(\hat{N_5}(\hat{N_5}+4)+9\right)\Psi^{abc}.$$
It is clear that this operator cannot lead to wave equations on
the cone since it is a constant. So intrinsic wave operators are
used to obtain wave equations on the cone. These operators exist
only in exceptional cases. For tensor fields of degree $
-1,0,1,...$, the intrinsic wave operators are $
\partial^2, (\partial^2)^2, (\partial^2)^3,...$ respectively
\cite{fr}. Thus, the following CI system of equations has been
utilized on the cone \cite{me}: \b \left\{ \ba{rcl}
(\partial_a\partial^a)^n \Psi&=&0,\\
\hat{N_5}\Psi&=&(n-2)\Psi.\ea\right. \e Other CI conditions can be
added to the above system in order to restrict the space of the
solutions. In order to project the coordinates on the cone $u^2=0$
to the dS space, we choose the following relation: \b \left\{
\ba{rcl}
x^{\alpha}&=&(Hu^5)^{-1}u^\alpha,\\
x^5&=&Hu^5.\ea\right.\e Note that $x^5$ becomes superfluous when
we deal with the projective cone. It is easy to show that various
intrinsic operators introduced previously now read as:
\begin{enumerate}

\item{ the conformal-degree operator $(\hat{N_5})$} \b \hat{N_5}=
 x_5\frac{\partial}{\partial x_5},\e

\item{the conformal gradient $(Grad_{\alpha})$}  \b Grad_{\alpha}=
-x_{5}^{-1}
\{H^2x_{\alpha}[Q_{0}-\hat{N_5}(\hat{N_5}-1)]+2\bar{\partial}_{\alpha}(\hat{N_5}+1)\},\e
where $\bar
\partial_\alpha$ is  tangential (or transverse) derivative in
de Sitter space $$ \bar
\partial_\alpha=\theta_{\alpha \beta}\partial^\beta=
\partial_\alpha  +H^2x_\alpha x\cdot\partial,\;\;\;x\cdot\bar
\partial=0.$$
$\theta_{\alpha \beta}=\eta_{\alpha \beta}+H^2x_{\alpha}x_{
\beta}$ is the transverse projector.
$Q_{0}=-{{1}\over{2}}M_{\alpha\beta}M^{\alpha\beta}=-H^{-2}(\bar\partial)^2$
is the scalar Casimir operator.

\item{and the powers of d'Alembertian $(\partial_a\partial^a)^n$},
which act intrinsically on field of conformal degree $(n-2)$, \b
(\partial_{a}
\partial^{a})^n=-H^{2n}x_{5}^{-2n} \prod_{j=1}^{n}[Q_{0}+(j+1)(j-2)]\,. \e
\end{enumerate}

We have shown \cite{me} that for scalar and vector fields, the
simplest CI system of equations is obtained through setting $n=1$
in $(2.5)$, {\it i.e.} the field with conformal degree $-1$.
Resulting field equations are transformed according to the UIRs of
$SO(1,4)$. In the flat limit $(H \rightarrow 0)$, the CI equation
for the vector field reduces exactly to the Maxwell equation
\cite{massless}. For a symmetric tensor field of rank-2, the CI
system $(2.5)$ with $n=1$ leads to \cite{derotata} (for simplicity
from now on we take $H=1$) : \b
(Q_0-2)\K_{\alpha\beta}+\frac{2}{3}{\cal
S}(\bar\partial_{\beta}+2x_{\beta})\bar\partial\cdot
\K_{\alpha}-\frac{1}{3}\theta_{\alpha\beta}\bar\partial\cdot\bar\partial\cdot\K=0\,.\e
By imposing the traceless and divergenceless conditions on the
tensor field ${\cal K}_{\alpha\beta}$, which are necessary for
UIRs of dS group, the CI equation $(2.10)$ reduces to
\footnote{\,\,$Q_2$ ($Q_1$) is the Casimir operator of the dS
group for the spin-2 (spin-1) field, mathematical details can be
found in \cite{massless,dix}.}
$$ (Q_0-2){\cal K}_{\alpha\beta}=0,\;\mbox{or}\,\,
\;(Q_{2}+4){\cal K}_{\alpha\beta}=0.$$ The solution of this CI
field equation corresponds to a representation of discrete series,
namely $\Pi^{\pm}_{2,1}$ \cite{dix,rotata}. However, this equation
does not coincide with any UIR of the Poincar\'e group In the flat
limit. Note that in the flat limit the CI equation $(2.10)$
reduces to the CI massless spin-2 wave equation of order-2 in four
dimensional Minkowski space which was found by Barut and Xu
\cite{derotata}; they have found this equation by varying the
coefficients of various terms in the standard equation
\cite{bx}.\\
If we take $n=2$ in $(2.5)$ we will obtain the following CI system
\cite{derotata}
$$ (Q_{2}+4)[(Q_{2}+6) {\cal
K}_{\alpha\beta}+D_{2\alpha}\partial_2 . {\cal K}_{\beta}]+ \frac
{1}{3}D_{2\alpha}D_{1\beta} \bar\partial . \bar \partial . {\cal
K}- \frac{1}{3} \theta_{\alpha\beta}(Q_{0}+6)\bar\partial . \bar
\partial . {\cal K}=0,$$ $$Q_{1}\bar
\partial . {\cal K}_{\alpha}+\frac{2}{3}D_{1\alpha}\bar \partial
.\bar\partial .{\cal K} +\frac{1}{6}Q_{1}D_{1\alpha}\bar \partial
.\bar\partial .{\cal K}=0,$$ \b \K'=0\,.\e By imposing the
traceless and divergenceless conditions on the tensor field ${\cal
K}_{\alpha\beta}$, the CI system $(2.11)$ becomes \b
 (Q_0-2)Q_0{\cal K}_{\alpha\beta}=0,\;\mbox{or}\,\,\; (Q_{2}+4)(Q_{2}+6) {\cal
K}_{\alpha\beta}=0.\e The solution of this CI field equation
corresponds to the two representations of discrete series, namely
$\Pi^{\pm}_{2,1}$ and $\Pi^{\pm}_{2,2}$ \cite{dix,rotata}. However
as mentioned, symmetric tensor field of rank-2 does not correspond
to any UIR of the conformal group. In the next section we will
study a mixed symmetry tensor.

\setcounter{equation}{0}
\section{Conformally invariant field equation}

Considering the conformal invariance in the dS space, we consider
a mixed symmetry rank-3 tensor field $\Psi_{abc}$ in Dirac's
null-cone. We classify the degrees of freedom of this tensor field
in de Sitter space by \footnote{we have used the notation
$x^\alpha x^\gamma\Psi_{\alpha\beta\gamma}\equiv
x\cdot\Psi_{\cdot\beta\cdot}$} \b
F_{\alpha\beta\gamma}=\Psi_{\alpha\beta\gamma}+x_\alpha
x\cdot\Psi_{\cdot\beta\gamma}+x_\beta
x\cdot\Psi_{\alpha\cdot\gamma}+x_\gamma
x\cdot\Psi_{\alpha\beta\cdot}+x_\alpha x_\gamma x\cdot
x\cdot\Psi_{\cdot\beta\cdot}+x_\beta x_\gamma x\cdot x\cdot
\Psi_{\alpha\cdot\cdot}
   ,\e   \b {\cal T}_{\alpha\beta}=
x\cdot\Psi_{\cdot\alpha\beta}+ x\cdot\Psi_{\alpha\cdot\beta}+
x\cdot\Psi_{\alpha\beta\cdot}+x_\alpha x\cdot x\cdot
\Psi_{\cdot\beta\cdot}+x_\beta  x\cdot x\cdot
\Psi_{\alpha\cdot\cdot},\e \b K_{\alpha}=  x\cdot x\cdot
\Psi_{\alpha\cdot\cdot},\e $$ \phi= x\cdot x\cdot x\cdot\Psi=0,$$
where $F_{\alpha\beta\gamma}$ is a mixed symmetry rank-3 tensor
field and $ {\cal T}_{\alpha\beta}$ and $K_{\alpha}$ are tensors
of rank two and one on dS space respectively $(x^\alpha {\cal
T}_{\alpha\beta}=x^\beta {\cal T}_{\alpha\beta}=0 = x^\alpha
K_\alpha)$. The fields $ \Psi_{5ab}$ and their contraction and
multiplication with $x$, are auxiliary fields and do not need to
be transformed to dS space. \\
We find CI equations in Dirac's null-cone by setting $n=2$ in
(2.5). Then followed by projection we obtain CI equations in dS
space  \b \left\{ \ba{rcl}
(Q_0-2)Q_0\Psi^{abc}&=&0,\\
\hat{N_5}\Psi^{abc}&=&0.\ea\right.\e \\
The following conditions can be added to the above system to
restrict the space of solutions:
\begin{enumerate}
\item[a)]{ transversality} $ u_a\Psi^{abc}=0 ,$ that results in \b
x^{5}(\Psi_{5bc}+x\cdot \Psi_{\cdot bc})=0,\e \item[b)]{
divergencelessness}  $Grad_a\Psi^{abc}=0,$ that results in \b
\partial \cdot \Psi_{\cdot\beta\gamma}=-x\cdot \partial
x\cdot\Psi_{\cdot\beta\gamma}, \; \mbox{or} \; \bar
\partial \cdot \Psi_{\cdot\beta\gamma}=-x\cdot
\Psi_{\cdot\beta\gamma}.\e
\end{enumerate}
We combine $(3.2)$, $(3.3)$ and $(3.1)$ to get : \b
Q_0(Q_0-2)x\cdot\Psi_{\cdot\beta\gamma}=0,\;\;Q_0(Q_0-2)x\cdot
x\cdot\Psi_{\alpha\cdot \cdot}=0,\;\; \bar\partial_\gamma
x\cdot\Psi_{\cdot\beta\gamma}=\Psi_{\gamma\beta\gamma}.\e  We can
write $F_{\alpha\beta\gamma}$ as \b \label{f}
F_{\alpha\beta\gamma}=\frac{1}{4}x_\gamma{\cal{A}}_{\alpha\beta} +
\Psi_{\alpha\beta\gamma}+ x_\alpha
x\cdot\Psi_{\cdot\beta\gamma}+x_\beta
x\cdot\Psi_{\alpha\cdot\gamma}\,,\e where we used the following
identities
$${\cal{A}}_{\alpha\beta}\equiv\bar\partial^\gamma
F_{\alpha\beta\gamma}-x_\alpha
F_{\gamma\beta}^{\;\;\;\;\gamma}+x_\beta
F_{\gamma\alpha}^{\;\;\;\;\gamma}=$$\b4(x\cdot\Psi_{\alpha\beta\cdot}+x_\alpha
x\cdot x\cdot\Psi_{\cdot\beta\cdot}+x_\beta x\cdot
x\cdot\Psi_{\alpha\cdot\cdot}),\e
$$\frac{1}{2}\bar\partial\cdot
F_{\alpha\cdot}^{\;\;\;\alpha}=x\cdot\Psi_{\alpha\cdot}^{\;\;\;\alpha},\,\,\,\,
F_{\alpha\beta}^{\;\;\;\;\alpha}-\frac{1}{2}x_\beta\bar\partial\cdot
F_{\alpha\cdot}^{\;\;\;\alpha}=\Psi_{\alpha\beta}^{\;\;\;\;\alpha}+x\cdot
x\cdot\Psi_{\cdot\beta\cdot }.$$ So the operation of $Q_0(Q_0-2)$
on $F_{\alpha\beta\gamma}$ leads to
$$Q_0(Q_0-2)(F_{\alpha\beta\gamma}-\frac{1}{4}x_\gamma{\cal{A}}_{\alpha\beta})=$$\b-4(\bar\partial_\alpha+3x_\alpha)
(Q_0-2)x\cdot\Psi_{\cdot\beta\gamma}-4(\bar\partial_\beta+3x_\beta)
(Q_0-2)x\cdot\Psi_{\alpha\cdot\gamma}.\e  Multiplying above
equation by $x_\beta$ results in \b
(Q_0-2)x\cdot\Psi_{\alpha\cdot\gamma}=\frac{1}{8}(Q_0-2)(4\bar\partial\cdot
F_{\alpha\cdot\gamma}-{\cal{A}}_{\alpha\gamma}-x_\gamma
\bar\partial\cdot{\cal{A}}_{\alpha\cdot}),\e similarly we have
\b(Q_0-2)x\cdot\Psi_{\cdot\beta\gamma}=\frac{1}{8}(Q_0-2)(4\bar\partial\cdot
F_{\cdot\beta\gamma}-{\cal{A}}_{\gamma\beta}-x_\gamma
\bar\partial\cdot{\cal{A}}_{\cdot\beta}).\e

Finally, from Eq.s $(3.7)$, $(3.8)$ and $(3.9)$, the following CI
field equation is obtained for the mixed symmetry tensor field
$F_{\alpha\beta\gamma}$ in de Sitter space $$
2Q_0(Q_0-2)(F_{\alpha\beta\gamma}-\frac{1}{4}x_\gamma{\cal{A}}_{\alpha\beta})+(\bar\partial_\alpha+3x_\alpha)
(Q_0-2)(4\bar\partial\cdot
F_{\cdot\beta\gamma}-{\cal{A}}_{\gamma\beta}-x_\gamma
\bar\partial\cdot{\cal{A}}_{\cdot\beta})$$\b\label{n}
+(\bar\partial_\beta+3x_\beta)(Q_0-2)(4\bar\partial\cdot
F_{\alpha\cdot\gamma}-{\cal{A}}_{\alpha\gamma}-x_\gamma
\bar\partial\cdot{\cal{A}}_{\alpha\cdot})=0.\e It is important to
note that the solution of this field equation is a physical state
of the conformal group which transforms according to the UIR of
this group. In the next section we will consider its
transformation according to UIRs of the de Sitter group $SO(1,4)$.

\setcounter{equation}{0}
\section{Group theoretical content}

In order to obtain the relation between the rank-3 mixed symmetry
tensor field $F_{\alpha\beta\gamma}$, and a massless spin-2 field
$\K_{\alpha\beta}$ (UIR of dS group), we define following
homomorphisms between them. There are different definitions. Here
we consider two cases.

\subsection{Simplest case}

The simplest homomorphism can be defined as: \b
F_{\alpha\beta\gamma}=\bar z_\alpha\K_{\beta\gamma}-\bar
z_\beta\K_{\alpha\gamma},\e where $\bar
z_\alpha=\theta_{\alpha\beta}z^\beta$ and $z^\beta$ is a constant
vector field. Replacing the equation $(4.1)$ in the field equation
$(3.10)$ and after some calculation, we obtain $$
Q_0(Q_0-2)\{12(x\cdot
z)\K_{\beta\gamma}+4(z\cdot\bar\partial)\K_{\beta\gamma}-5\bar
z_\beta \bar\partial . \K_\gamma-4x_\beta z.\K_\gamma-\bar
z_\gamma \bar\partial . \K_\beta$$\b-3x_\gamma (x.z)
\bar\partial.\K_\beta-x_\gamma
(z.\bar\partial)\bar\partial.\K_\beta+x_\beta x_\gamma
z.\bar\partial.\K+x_\gamma\bar
z_\beta\bar\partial.\bar\partial.\K\}=0.\e By imposing the
traceless and divergenceless conditions which are necessary for
associating $\K_{\alpha\beta}$ with the UIR of the dS group,  we
get \b 7[Q_0]^2\K_{\beta\gamma}-46 Q_0\K_{\beta\gamma}- 64
\K_{\beta\gamma}=0,\; \mbox{or}\, \;
(Q_0-2)(7Q_0-32)\K_{\beta\gamma}=0.\e Clearly this equation is not
compatible with equation $(2.12)$ of the de Sitter linear gravity.
In other words this homomorphism cannot lead to any UIR of dS
group. Now we consider another possibility.

\subsection{Second case}

Now we try the following definition, which is deduced from the
field strength tensor of electromagnetic potential: \b
F_{\alpha\beta\gamma} \equiv\Big(\bar
\partial_\alpha +x_\alpha\Big)\K_{\beta\gamma}-\Big(\bar
\partial_\beta +x_\beta
\Big)\K_{\alpha\gamma}.\e  By substituting $(4.4)$ into the
$(3.10)$, we find
$$Q_0(Q_0-2)\Big[4(Q_0-2)\K_{\beta\gamma}-Q_0x_\gamma\bar\partial\cdot\K_\beta+
3\bar\partial_\beta\bar\partial\cdot\K_\gamma+7
x_\beta\bar\partial\cdot\K_\gamma-x_\gamma\bar\partial\cdot\K_\beta$$\b-\bar\partial_\gamma
\bar\partial\cdot\K_\beta-2x_\beta
x_\gamma\bar\partial\cdot\bar\partial\cdot\K
-x_\gamma\bar\partial_\beta\bar\partial\cdot\bar\partial\cdot\K\Big]=0.\e
Similarly by imposing the traceless and divergenceless conditions
($\K'=0=\bar
\partial \cdot \K$), we obtain \b (Q_0-2)^2Q_0\K_{\beta\gamma}=0,\,\,\,\mbox{or equivalently},\,\,
(Q_{2}+4)^2(Q_{2}+6){\cal K}_{\alpha\beta}=0.\e It is clear that
this CI field corresponds to the two representations of discrete
series, namely $\Pi^{\pm}_{2,1}$ (twofold) and $\Pi^{\pm}_{2,2}$.
The representations $\Pi^{\pm}_{2,2}$ of the discrete series have
a Minkowskian interpretation and have a unique extension to a
direct sum of two UIRs $C(3;2,0)$ and $C(-3;2,0)$ of the conformal
group with positive and negative energies, respectively \cite{ms9,
ms7}. $\Pi^{+}_{2,2}$ restricts to the massless UIR ${\cal P}^>(0,
2)$ (${\cal P}^<(0,2)$) of the Poincar\'e group with positive
(negative) energy. Similar statements hold for $\Pi^{-}_{2,2}$
with negative helicity (namely $ {\cal P}^{ \stackrel{>}
{<}}(0,-2)$). Moreover, equation (4.6) can be written in the
intrinsic coordinates as \cite{derotata,rotata}: \b \left(\Box^3+8
\Box^2+24 \Box+48 \right)h_{\mu\nu}=0,\e and in the metric
signature $(-,+,+,+)$, we have: \b \left(\Box^3-8 \Box^2+24\Box-48
\right)h_{\mu\nu}=0.\e Therefore if one insists $\K_{\alpha\beta}$
or equivalently $h_{\mu\nu}$ to transform according to the UIR of
dS and conformal groups it must satisfy a field equation of order
6.

\subsection{Fierz representation}

The spin-2 field can be described in two ways, which are called
the Einstein frame and the Fierz frame representations. The most
common one, the Einstein frame, uses a symmetric tensor of rank-2,
$\K$ to represent the field. In the Fierz frame this role is
played by a mixed symmetry tensor of rank-3, $F$. Such an object
has 20 independent components. It has been shown that it must obey
a further condition in order to represent only one single spin-2
field, otherwise it represents two spin-2 fields \cite{frhe,none}.
\\Now we try the Fierz representation. Transformation of the Fierz
representation from intrinsic coordinate to the ambient space
results in \cite{derotata,rotata}:
$$ F_{\alpha\beta\gamma} \equiv\Big(\bar
\partial_\alpha +x_\alpha\Big)\K_{\beta\gamma}-\Big(\bar
\partial_\beta +x_\beta
\Big)\K_{\alpha\gamma}$$ \b
+\theta_{\beta\gamma}\Big(\bar\partial_\alpha
\K-\bar\partial\cdot\K_{\alpha}-x_\alpha\K\Big)-\theta_{\alpha\gamma}\Big(\bar\partial_\beta
\K-\bar\partial\cdot\K_{\beta}-x_\beta\K\Big).\e It is interesting
to note that by imposing the traceless and divergenceless
conditions, which are necessary for associating with UIRs, Fierz
representation reduces to the previous case, and one can associate
the same UIR of the dS group. The difference between these two
cases is that the latter is an isomorophism between the mixed
symmetry rank-3 tensor field $F$ and symmetric rank-2 tensor field
$\K$.

\section{Conclusion}
It was pointed out that Einstein's theory of gravitation, in the
background field method, $g_{\mu\nu}=g_{\mu\nu}^{BG}+h_{\mu\nu}$,
can be considered as a theory of massless symmetric tensor field
of rank-2 on a fixed background, such as dS space. Massless fields
propagate on the light cone and then their equations must be CI.
Contrary to Maxwell equation, Einstein's equation of gravitation,
as well as equation of $h_{\mu\nu}$, is not conformally invariant.
 \\In our previous paper
\cite{derotata} we used a symmetric rank-2 tensor field
$\Psi_{ab}$ and Dirac's six-cone formalism to obtain CI field
equation for $\K_{\alpha\beta}$ in dS space. Although the equation
was CI, it did not transform according to the UIRs of the
conformal group i.e. it was not physical state of this group.
Binegar et al \cite{co5} have shown that mixed symmetry tensor
field of rank-3 transforms according to UIRs of the conformal
group. In this paper, by definition homomorphisms between this
tensor field and $\K_{\alpha\beta}$, we obtained a CI equation
which can be interpreted as the UIR of the conformal and dS
groups. It has been shown that if we want $\K_{\alpha\beta}$ to be
a physical state of the dS and conformal groups simultaneously, it
must satisfy a field equation of order 6. So, conformal gravity
seems to be like a $R^3$ gravity theory. As a future work, it may
be possible to find a CI gravitational field which in its linear
approximation gives this linear physical equation.

\vskip 0.5 cm

\noindent {\bf{Acknowledgement}}:  The authors would like to thank
Prof. M.B. Paranjape.


\begin{thebibliography}{a}
\addcontentsline{toc}{chapter}{Bibliographie}

\bibitem{co5} B. Binegar, C. Fronsdal and W. Heidenreich, Phys. Rev. D 27, 2249 (1983).
\bibitem{ms9} A.O. Barut, A. B\"ohm, J. Math. Phys. 11, 2938 (1970).
\bibitem{me} S. Behroozi, S. Rouhani, M.V. Takook and M.R. Tanhayi, Phys. Rev. D 74, 124014 (2006).
\bibitem{derotata} M. Dehghani, S. Rouhani, M.V. Takook and M.R. Tanhayi,
Phys. Rev. D 77, 064028 (2008).
\bibitem{dir} P. A. M. Dirac, Ann. of Math. 36, 657 (1935); 37, 429 (1935-b).
\bibitem{s6} G. Mack and A. Salam, Ann. Phys. 53, 174 (1969).
\bibitem{o7} H.A. Kastrup, Phys. Rev. 150, 1189 (1964); S.L. Adler
Phys. Rev. D 6, 3445 (1972); C.R. Preitschop, M.A. Vosiliev,
hep-th/9812113; P. Arvidsson, JHEP 03, 076 (2006).
\bibitem{frhe} C. Fronsdal, W. Heidenreich, J. Math. Phys. 28, 215 (1987).
\bibitem{none} M. Novello, R.P. Neves, Class. Quant. Grav. 19, 5335 (2002).
\bibitem{ja} A.G.Riess et al. [Supernova Search Team Collaboration], Astro. J. 116,
1009 (1998); S.Perlmutter et al. [Supernova Cosmology Project
Collaboration], Astro. J. 517, 567 (1999); U. Seljak, A. Slosar,
and P. McDonald, JCAP, 014, 610 (2006); A.G. Riess et al., Astro.
J. 98, 659 (2007).
\bibitem{ja2} A.D. Linde, \emph{Particle Physics and Inflationary Cosmology}, (Harwood Academic Publishers,
Chur, Switzerland 1990).
\bibitem{fr}  C. Fronsdal, Phys. Rev. D 20, 848 (1979).
\bibitem{massless} T. Garidi, J.P. Gazeau, S. Rouhani and M.V. Takook, J. Math. Phys. 49, 032501 (2008);
T. Garidi, J.P. Gazeau and M.V. Takook, J. Math. Phys. 44, 3838
(2003).
\bibitem{dix} J. Dixmier, Bull Soc. Math. France 89, 9 (1961); B. Takahashi, Bull. Soc. Math. France 91, 289 (1963).
\bibitem{rotata} S. Rouhani, M.V. Takook and M.R. Tanhayi, {\it Linear Weyl Gravity in de Sitter space},
arXive:0903.2670.
\bibitem{bx} A.O. Barut and B.W. Xu, J. Phys. A 15, L207 (1982).
\bibitem{ms7} M. Levy-Nahas, J. Math.
Phys. 8, 1211 (1967).

\end{thebibliography}
\end{document}